\documentclass[10pt,twocolumn,floatfix,superscriptaddress,preprintnumbers,amsmath,amssymb,prb]{revtex4-2}
\usepackage{graphicx}
\usepackage{dcolumn}
\usepackage{bm}
\usepackage{amssymb,amsmath}
\usepackage{float}
\usepackage{color}
\usepackage{hyperref}
\usepackage{ulem}
\begin{document}

\preprint{preprint(\today)}

\title{Unconventional pressure dependence of the superfluid density in the nodeless  topological superconductor $\alpha$--PdBi$_2$}

\author{Debarchan Das}
\email{debarchandas.phy@gmail.com}
\affiliation{Laboratory for Muon Spin Spectroscopy, Paul Scherrer Institute, CH-5232 Villigen PSI, Switzerland}
\author{Ritu Gupta}
\affiliation{Laboratory for Muon Spin Spectroscopy, Paul Scherrer Institute, CH-5232 Villigen PSI, Switzerland}
\author{Christopher Baines}
\affiliation{Laboratory for Muon Spin Spectroscopy, Paul Scherrer Institute, CH-5232 Villigen PSI, Switzerland}
\author{Hubertus Luetkens}
\affiliation{Laboratory for Muon Spin Spectroscopy, Paul Scherrer Institute, CH-5232 Villigen PSI, Switzerland}
\author{Dariusz Kaczorowski}
\affiliation{Institute of Low Temperature and Structure Research, Polish Academy of Sciences, Wroc\l aw, ul. Ok\'olna 2, 50-422, Poland}
\author{Zurab Guguchia}
\affiliation{Laboratory for Muon Spin Spectroscopy, Paul Scherrer Institute, CH-5232 Villigen PSI, Switzerland}
\author{Rustem Khasanov}
\email{rustem.khasanov@psi.ch}
\affiliation{Laboratory for Muon Spin Spectroscopy, Paul Scherrer Institute, CH-5232 Villigen PSI, Switzerland}

\begin{abstract}
We investigated the superconducting properties of the topological superconductor $\alpha$--PdBi$_2$ at ambient and external pressures up to 1.77~GPa  using muon spin rotation ($\mu$SR) experiments. The ambient pressure measurements evince a fully gapped $s$-wave superconducting state in the bulk of the specimen.
AC magnetic susceptibility and $\mu$SR measurements manifest a continuous suppression of $T_{\rm c}$ with increasing pressure. In parallel, we observed a significant decrease of superfluid density by $\sim$20\% upon application of external pressure. Remarkably, the superfluid density follows linear relation with $T_{\rm c}$ which was found before in some unconventional topological superconductors and hole doped cuprates. This finding signals a possible crossover from BEC to BCS in $\alpha$--PdBi$_2$. 


\end{abstract}


\maketitle


Topological superconductors (TSCs) possess a full pairing gap in the bulk and gapless surface states protected by time-reversal symmetry (TRS) \cite{Qi,Ando, Sato, Hasan}. The general interest to search for novel TSC is associated with its intimate connection to Majorana bound states as low energy excitation \cite{Hasan,Wilczek} which have potential applications in quantum computation \cite{Kitaev, Ivanov}. Thus, TSC has attracted a significant research interest in recent time due to its scientific importance and potential applications prospect. Despite serious research efforts, the experimental realizations of bulk topological superconductivity in real materials remain considerably limited to few classes of materials, such as Cu/Sr/Nb-doped Bi$_2$Se$_3$ \cite{Hor, Kriener, Smylie1, Kobayashi, Krieger, Das}, layered transition metal dichalcogenides \cite{GuguchiaMoTe2,Guguchia2,Rohr, Noh}, ternary transition metal pnictides \cite{Qian, Das2}, In-doped SnTe \cite{Sasaki} and few half-Heusler compounds \cite{Chadov, Lin}. Interestingly, recent reports investigating various members of  Pd-Bi family reinvigorate the interest to search for potential TSCs in this class of materials. Due to their intrinsic capability to maintain strong spin orbit coupling (SOC), Pd-Bi family turns out to be quite promising. For instance, noncentrosymmetric $\alpha$-PdBi with a superconducting transition at $T_{\rm c}$ = 3.8~K \cite{Joshi} exhibits Dirac cone electronic dispersion at 700~meV \cite{Sun, Neupane} below the Fermi energy. Similarly, the compound $\beta$-PdBi$_2$ is another member of the Pd-Bi family which has been reported to possibly host topological superconducting state below $T_{\rm c}$ = 5.3~K~\cite{Imai, Biswas, Kacma}. The Dirac point in this material is located at 2.41~eV below the Fermi energy \cite{Sakano, Iwaya, Kolapo, Chen}. In this framework, the relatively new member of Pd-Bi family, $\alpha$-PdBi$_2$, crystallizing  in monoclinic centrosymmetric structure (space group $C2/m$), turns out to be very promising due to its superconducting ($T_{\rm c}$  = 1.7~K) as well as nontrivial topological properties \cite{Mitra, Choi, Dimitri}.

Recent ARPES results \cite{Dimitri} on this compound reveals the presence of a Dirac point at 1.26~eV below the Fermi energy at the zone center. Notably, the ARPES data also display multiple band crossings at the Fermi energy. This tempts us to speculate that the superconducting gap structure might consist of multiple gaps.  However, the tunnel diode oscillator (TDO) based penetration depth study \cite{Mitra} indicates an isotropic gap structure in this compound with the gap to $T_{\rm c}$ ratio, $\Delta(0)/k_{\rm B}T_{\rm c}$ $\sim$~2.0. However, it must be noted that TDO studies are very sensitive to surface superconductivity. Therefore, it is desired to probe the superconducting gap structure by means of a bulk experimental technique. Further, Zhou $et~al$.\cite{Zhou} recently reported a very interesting pressure ($p$)- $T_{\rm c}$ phase diagram  in this compound manifesting a sudden increase in $T_{\rm c}$ at $p$= 1.3~GPa. Interestingly, X-ray studies under pressure reveals no structural change around this pressure. Thus, it is very important to understand the origin of the rise in $T_{\rm c}$ and to check whether the superfluid density also shows such an enhancement with pressure. In this regard, muon spin rotation ($\mu$SR) experiments at ambient and as well as under different hydrostatic pressures are of great importance to address these aforementioned issues.
\begin{figure}[htb!]
\includegraphics[width=\columnwidth]{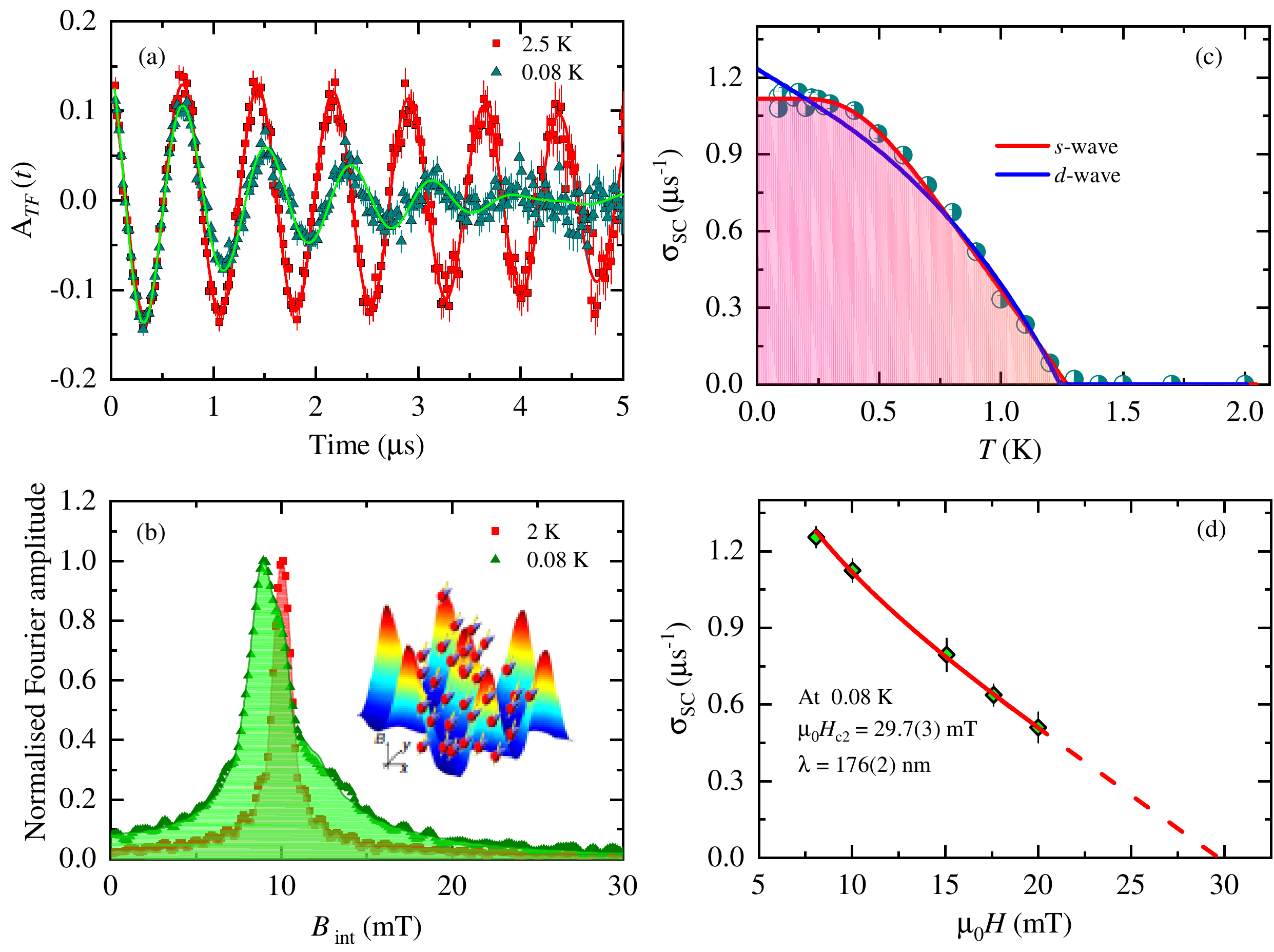}
\caption{(Color online) (a) Transverse-field (TF) ${\mu}$SR asymmetry spectra obtained above and below $T_{\rm c}$ for $\alpha$-PdBi$_2$. The spectra were recorded after field cooling the sample from above $T_{\rm c}$ in an applied magnetic field of ${\mu}_{\rm 0}H = 10$~mT . Solid lines correspond to the fitting of the spectra as described in the Supplemental Material (SM)\cite{supplemental}. (b) Normalised Fourier spectra at 0.08~K (green) and 2~K (red)  obtained by fast Fourier transformation of the spectra in (a). Inset: schematic illustration of how muons, as local probes, sense the inhomogeneous magnetic field distribution in the vortex state of a type-II superconductor. (c) Temperature dependence of the  superconducting muon spin depolarization rate ${\sigma}_{\rm sc}$ of $\alpha$-PdBi$_2$ measured in an applied magnetic field of ${\mu}_{\rm 0}H = 10$~mT. The solid lines correspond to different theoretical models as discussed in the text. (d) Field dependence of ${\sigma}_{\rm sc}$  fitted with the Brandt formula for an isotropic single $s$ wave gap (solid line) as discussed in the text.}
\label{fig1}
\end{figure}

In this letter, we report the results of our detailed $\mu$SR experiments on $\alpha$-PdBi$_2$ at ambient and under different hydrostatic pressures. Ambient pressure measurements substantiate single gap $s$-wave superconductivity with preserved TRS state in this compound. Furthermore, by performing AC-magnetic susceptibility and $\mu$SR experiments under different hydrostatic pressures, we explored the pressure- $T_{\rm c}$ phase diagram which is noticeably different from the previous report by Zhou $et~al$.\cite{Zhou}. Our results manifest a subtantial and continious decrease of superfluid density ($\sim$20\%) and $T_{\rm c}$ with application of pressure up to 1.77~GPa.

Pieces consisting of many single crystals (not cleaved) of $\alpha$-PdBi$_2$ obtained using melt growth technique \cite{Mitra} were used for the present study. Ambient pressure and high pressure TF ${\mu}$SR experiments were performed on HAL-9500 and GPD spectrometers at the Paul Scherrer Institute (Villigen, Switzerland) respectively. For pressure dependent measurements, we used, a double-wall piston cylinder type of cell with inner channel made of CuBe material, especially designed to perform ${\mu}$SR experiments under pressure \cite{Andreica,Khasanov,Shermadini} and Daphne oil (7373) as a pressure-transmitting medium. All Transverse-field (TF) experiments were done after a field-cooling procedure. The ${\mu}$SR asymmetry spectra were analyzed using the open software package MUSRFIT \cite{Suter}.




Figure~\ref{fig1}a depicts TF-$\mu$SR asymmetry spectra for $\alpha$-PdBi$_2$ at two selected temperatures above (2.5~K) and below (0.08~K) $T_{\rm c}$, measured in an applied magnetic field of 10~mT. Above $T_{\rm c}$, the spectra exhibits a very small relaxation due to the presence of random local fields associated with the nuclear magnetic moments. On the contrary, in the superconducting state, the formation of flux line lattice (FLL) produces an inhomogeneous distribution of magnetic field (inset of Fig.~\ref{fig1}b) which increases the relaxation rate of the $\mu$SR signal. Figure~\ref{fig1}b shows the normalised Fourier spectra for single-crystal $\alpha$-PdBi$_2$ at temperatures above (2.5~K) and below (0.08~K) $T_{\rm c}$ obtained by means of the maximum entropy Fourier transform technique. While we observed a sharp peak in the Fourier amplitude around 10~mT (external applied field) at 2.5~K confirming homogeneous field distribution throughout the sample, a fairly asymmetric broad signal was seen at 0.08~K evincing the fact that the sample is indeed in the superconducting mixed state. The formation of the FLL causes this broadening of the line shape.

The temperature dependence of ${\sigma}_{\rm sc}(T)$ for $\alpha$-PdBi$_2$ measured at an applied field of ${\mu}_{\rm 0}H=10$~mT is presented in Fig.~\ref{fig1}c. From the observed temperature dependence of ${\sigma}_{\rm sc}(T)$, the nature of the superconducting gap can be determined as ${\sigma}_{\rm sc}(T)$  is proportional to the superfluid density.
The solid red and blue lines in Fig.~\ref{fig1}c represent fitting of the experimental data with isotropic $s$-wave and $d$-wave models \cite{Tinkham,Prozorov,carrington}, respectively(see SM \cite{supplemental}). It is quite evident from the figure that the observed ${\sigma}_{\rm sc}(T)$ excludes presence of any nodes and it can be best described using single isotropic gap ($s$-wave) yielding a gap value of $\Delta(0)$ = 0.18(1)~meV and $T_{\rm c}= 1.27(3)$.   In this context, it is to be noted that the $T_{\rm c}$ obtained from ${\mu}$SR measurements is slightly lower than that determined from heat capacity measurements. This is because of the fact that ${\mu}$SR measurements were carried out in an applied field of 10~mT which slightly lowered the superconducting transition temperature. From $\mu$SR measurements we obtained the superconducting gap to $T_{\rm c}$ ratio as  $\Delta(0)/k_{\rm B}T_{\rm c}$ = 1.64(1), which is very close to the value expected (1.764) for BCS superconductors. It is worthwhile to emphasize that previous ARPES and band structure calculations \cite{Dimitri, Choi} evince the presence of a Rashba state near the Fermi energy in this compound. It is theoretically predicted that in presence of an in-plane magnetic field, the Majorana zero mode can be realized utilizing the coupling of an $s$-wave superconductor with a material exhibiting Rashba states \cite{Potter}. Thus, the  observation of $s$-wave superconductivity in $\alpha$-PdBi$_2$ is quite crucial in search for Majorana fermions.

\begin{figure}
\includegraphics[width=\columnwidth]{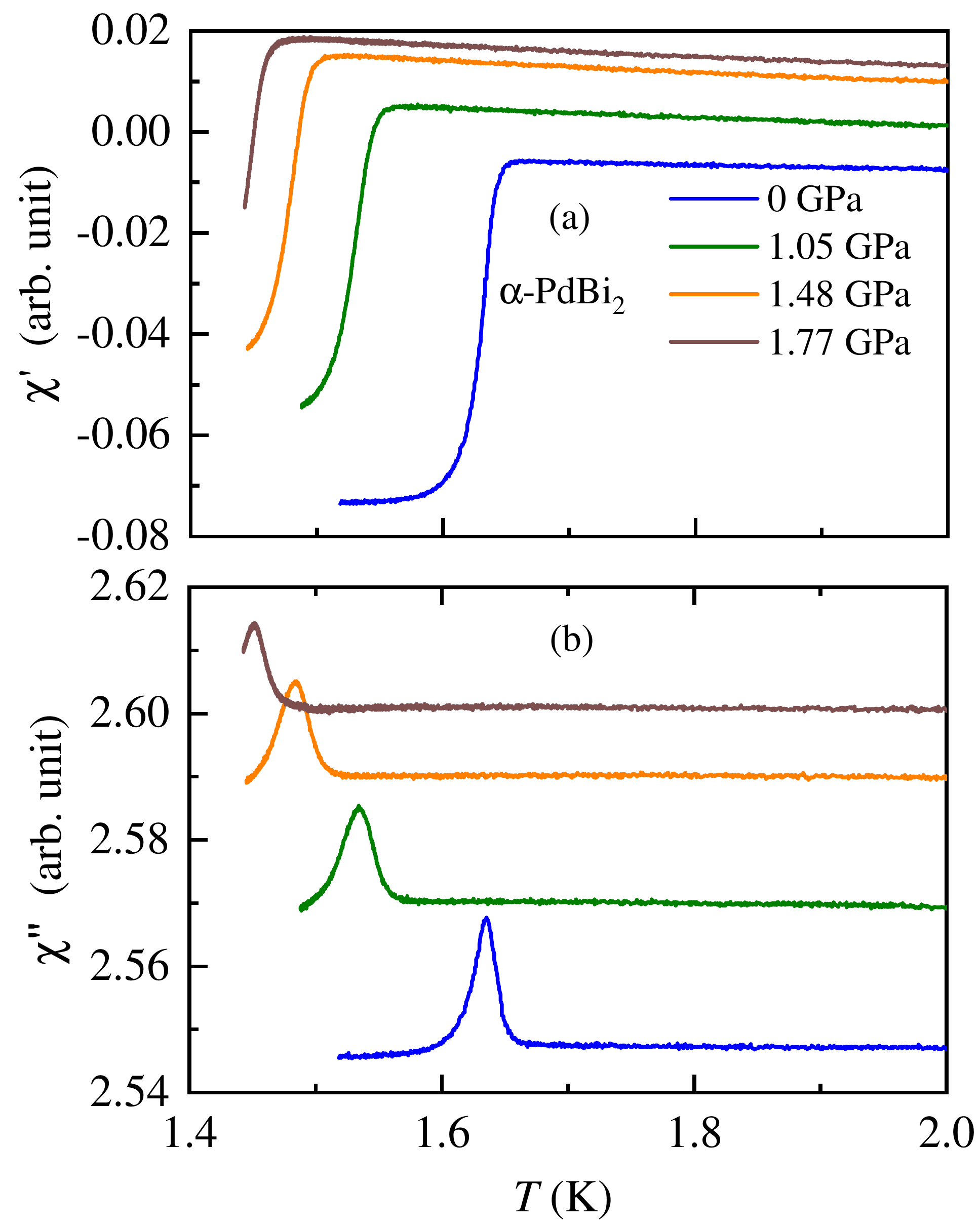}
\caption{(Color online) Temperature dependence of the real, $\chi^{\prime}$, (a) and imaginary, $\chi^{\prime\prime}$, (b) components of ACS of $\alpha$-PdBi$_2$  (mounted inside a double wall  piston-cylinder type pressure cell made of CuBe/MP35N material) at different selected pressures. For better visualization, we vertically shifted the y-axis in both (a) and (b).}
\label{fig2}
\end{figure}

Furthermore, we also studied the field dependence of TF-relaxation rate ${\sigma}_{\rm sc}(\rm B)$. Figure~\ref{fig1}d illustrates the field dependence of ${\sigma}_{\rm sc}(\rm B)$ at 0.08~K. At each point, data was taken after field cooling the sample from 2~K. We analyzed ${\sigma}_{\rm sc}(\rm B)$ using the Brandt formula \cite{Brandt} for an single $s$-wave.
Fitting with this model is shown by the red solid line in Fig.~\ref{fig1}d. From this analysis, we determined the values of upper critical field ($\mu_0H_{\rm c2}(0)$)  and effective London penetration depth ($\lambda_{\rm eff}$) as  29.7(3)~mT and 176(2)~nm, respectively. Thus obtained value of  $\mu_0H_{\rm c2}(0)$ and $\lambda_{\rm eff}$ are in reasonably good agreement with the previous reports \cite{Zhou,Mitra}. $\lambda_{\rm eff}$ is related to the superconducting carrier density $n_{\rm s}$ via the relation $\lambda^2_{\rm eff}$=$\left( m^*/\mu_0n_{\rm s}e^2\right)$, where $m^*$ is the effective mass. Now $m^*$ can be estimated from the relation $m^*=(1+\lambda_{\rm e-ph})m_{\rm e}$ where $m_{\rm e}$ is the electron rest mass and $\lambda_{\rm e-ph}$ is the electron–phonon coupling constant found out to be 0.54 from heat capacity (see SM \cite{supplemental}). Thus, we estimated $n_{\rm s}$~=~1.41$\times 10^{27}$~m$^{-3}$. This is comparable to the ones seen in some other TSCs, e.g. Nb$_{0.25}$Bi$_2$Se$_3$  ($n_{\rm s} = 0.25\times 10^{26}~$m$^{-3}$)~\cite{Das}, $T_d$-MoTe$_2$ ($n_{\rm s} = 1.67\times 10^{26}~$m$^{-3}$)~\cite{GuguchiaMoTe2} and 2M-WS$_2$ ($n_{\rm s} = 2.8\times 10^{26}~$m$^{-3}$)~\cite{Guguchia2}. Hall conductivity measurements estimating the carrier density in the normal state of this compound will be quite informative as it will  allow us to compare the carrier densities in the normal and superconducting states. This will also provide information on whether the single-gap superconductivity in $\alpha$-PdBi$_2$ originates from the superconducting gap occurring only on the electron or hole-like Fermi surface. Moreover, results of our zero-field ${\mu}$SR measurements (not shown) confirm that the TRS state is preserved in the superconducting state of $\alpha$-PdBi$_2$.
\begin{figure}[htb!]
\includegraphics[width=\columnwidth]{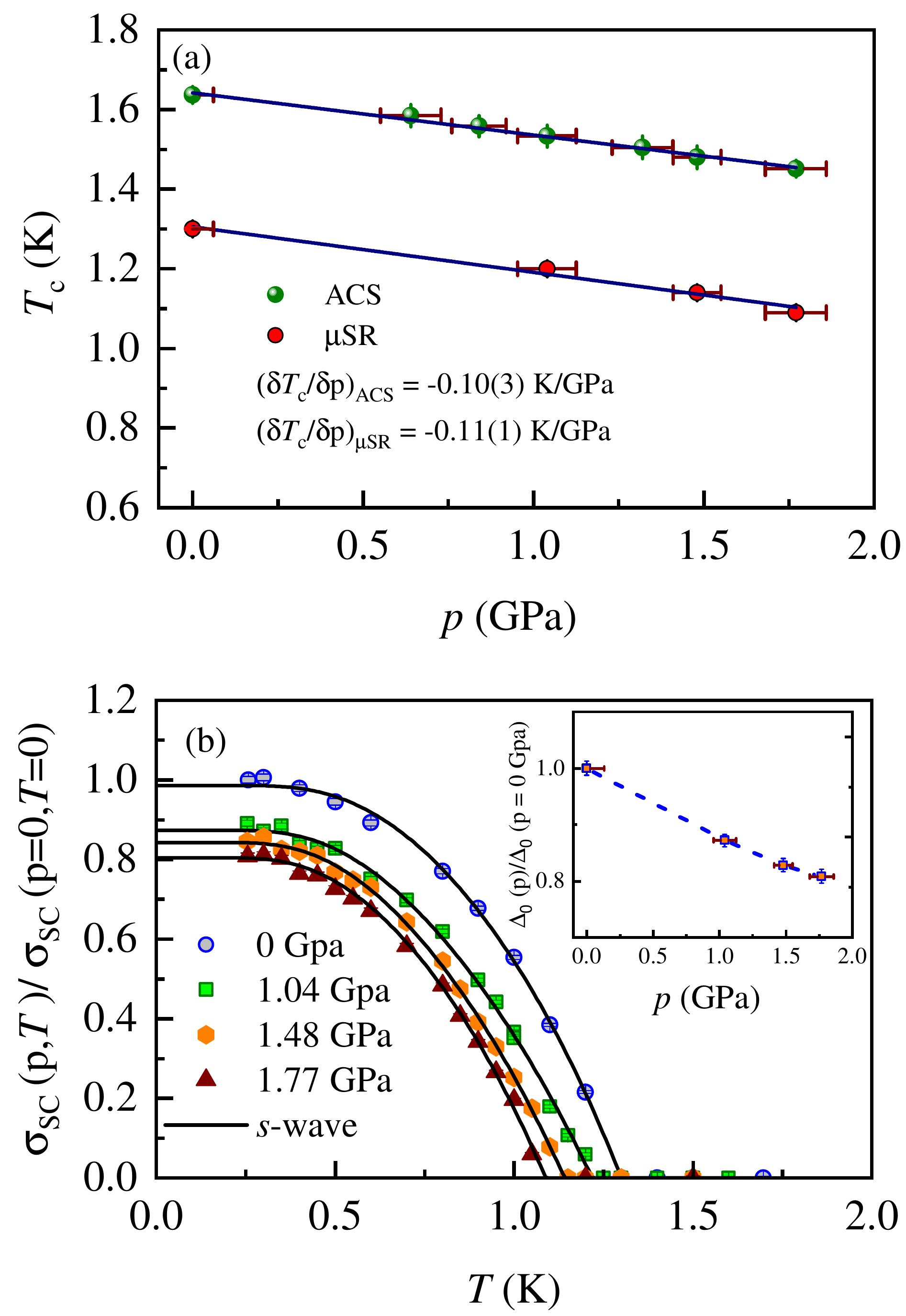}
\caption{(Color online) (a) Pressure($p$)-$T_{\rm c}$ phase diagram of $\alpha$-PdBi$_2$ constructed from ACS (green) and $\mu$SR (red) measurements under various applied hydrostatic pressures. Solid lines represent the linear fit through the experimentally obtained data points. Note that $T_{\rm c}$'s from $\mu$SR measurement are lower than that obtained from ACS because $\mu$SR experiments we performed under an applied field of 10~mT. (b) The temperature dependence of ${\sigma}_{\rm sc} (T)$ (normalized to its value at the zero applied pressure), under different applied pressures. Inset shows the pressure dependence of the superconducting gap $\Delta_0$ normalized to the zero pressure value.}
\label{fig3}
\end{figure}



Figure\ref{fig2}(a) and (b) show the real ($\chi^{\prime}$) and imaginary ($\chi^{\prime\prime}$) components of the magnetic ac-susceptibility (ACS) of $\alpha$-PdBi$_2$. The data in Fig.~\ref{fig2}(a) and (b) are shifted vertically for better visualization. In the superconducting state, we clearly see a clear drop in $\chi^{\prime}$ and a pronounced peak in $\chi^{\prime\prime}$. For constructing the pressure ($p$)-$T_{\rm c}$ phase diagram from ACS measurements (see Fig.~\ref{fig3}a), we considered the position of the peak in $\chi^{\prime\prime}$ (Fig.~\ref{fig2}b) as $T_{\rm c}$. It is quite evident that up to 1.77~GPa, $T_{\rm c}$ decreases with increasing pressure.  A linear fit to $T_{\rm c}(p)$ (blue solid line in Fig.~\ref{fig3}a) yields a slope $\frac{\delta T_{\rm c}}{\delta p}|_{\rm ACS}= -0.10(3)$~K/GPa. This observation is in contrast to that previously reported by Zhou $et~al.$ \cite{Zhou} where $T_{\rm c}$ showed a sudden jump with increasing pressure. It is important to note that for their transport measurements, Zhou $et~al.$ used NaCl powder as the pressure transmitting medium which does not guarantee proper hydrostatic condition. We speculate that such non-hydrostatic conditions can generate unexpectedly high pressures in micro grains of the sample resulting in local structural transition to $\beta$-PdBi$_2$ phase. This might have increased the $T_{\rm c}$ locally. As electrical transport measurements are sensitive to the filamentary superconducting channels, the enhancement in $T_{\rm c}$ with increasing pressure as reported by Zhou $et~al.$ \cite{Zhou} might be an artifact of non-hydrostatic condition. On the other hand, as discussed the SM \cite{supplemental}, all of our measurements were performed under hydrostatic conditions.

TF-$\mu$SR asymmetry spectra obtained from GPD experiments were analysed as per the formalism presented in ref.~ \cite{Shermadini}. In Fig.~\ref{fig3}b, we present the temperature dependence of ${\sigma}_{\rm sc}(p,T)$ under various hydrostatic pressures, after normalising it to the zero temperature value at ambient pressure $(i.e.~{\sigma}_{\rm sc} (p=0,T=0))$. As shown by the solid lines in  Fig.~\ref{fig3}b, a single gap $s$-wave model describes the ${\sigma}_{\rm sc}(T)$ data very well up to the highest pressure investigated. The pressure evolution of $T_{\rm c}$ determined from $\mu$SR experiments is shown in Fig~\ref{fig3}a. We note that the slope $\frac{\delta T_{\rm c}}{\delta p}|_{\mu \rm SR}= -0.11(1)$~K/GPa is very close to that determined from ACS measurements. Inset of Fig~\ref{fig3}a shows the pressure variation of the gap value normalised to its zero pressure value, $\Delta_0(p)/\Delta_0(p=0)$, which also decrease with increasing pressure.

\begin{figure}[htb!]
\includegraphics[width=\columnwidth]{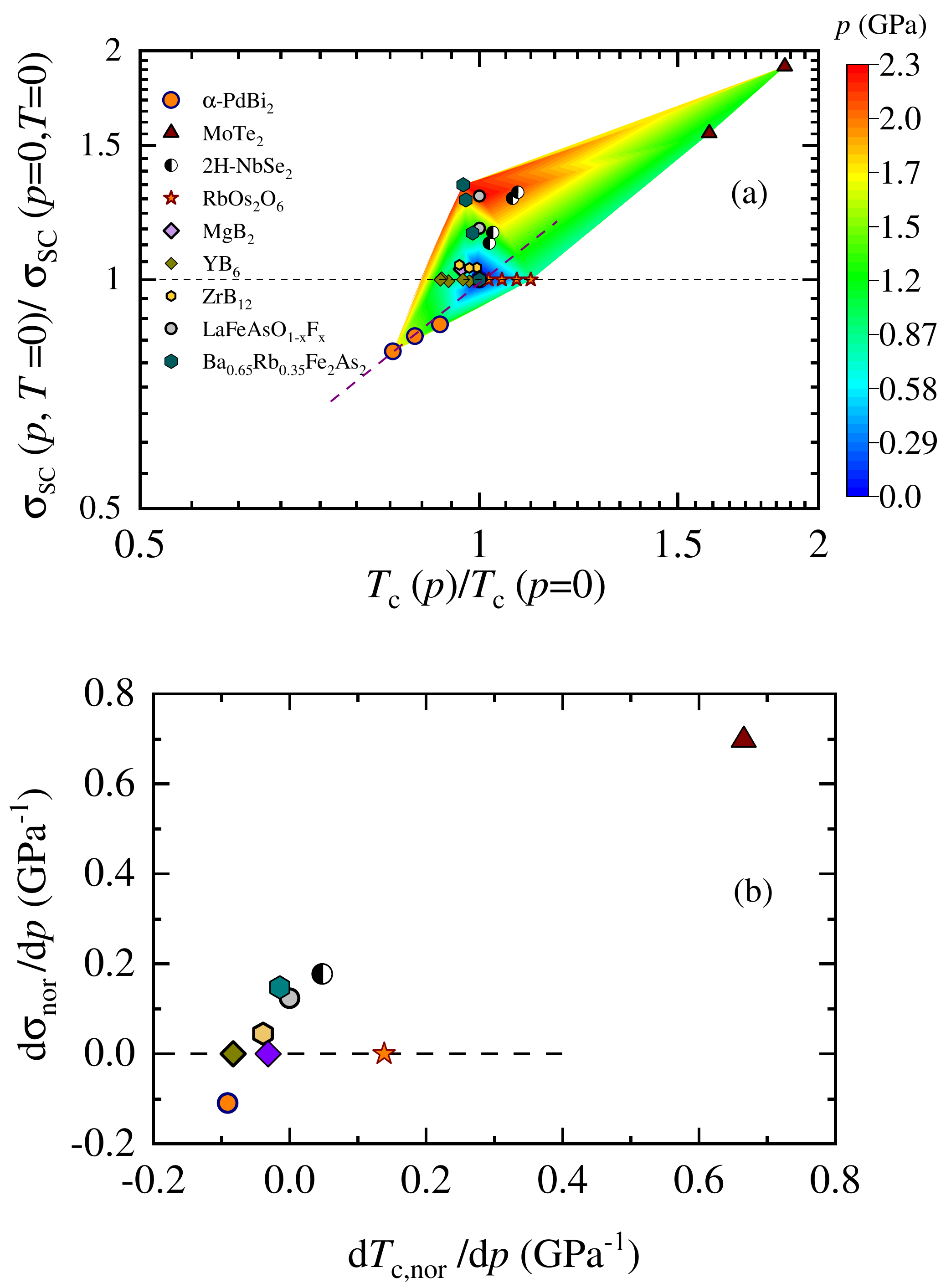}
\caption{(Color online)(a) Normalised superfluid density as a function of $T_{\rm c}(p)/T_{\rm c}(p=0)$ at different applied pressures shown in a color contour plot. We compared the pressure effect with few other conventional and unconventional superconductors. Black dashed horizontal line manifests BCS prediction. Purple dashed line indicates the deviation from BCS prediction and highlights linear dependence between superfluid density and $T_{\rm c}$ in $\alpha$-PdBi$_2$. (b) d$\sigma_{nor}$/d$p$ as a function of d$T_{\rm c, nor}$/d$p$ (where $X_{nor}$ =$\frac{X_p-X_{p=0}}{X_{p=0}}$, with $X$ being $T_{\rm c}$ or $\sigma_{\rm sc}$ for different superconductors depicted in panel (a).}
\label{fig4}
\end{figure}

Remarkably, the ratio $\frac{{\sigma}_{\rm sc}(p,T=0)}{{\sigma}_{\rm sc}(p=0,T=0)}$, which is a direct measure of relative change of the superfluid density, is decreased considerably by $\sim$20\% at 1.77~GPa. Such a significant reduction in superfluid density clearly indicates unconventional superconductivity in $\alpha$-PdBi$_2$  as in conventional BCS superconductors, the superfluid density either exhibits very weak pressure dependency or remains independent of pressure as highlighted by the horizontal dashed line in Fig.~\ref{fig4}(a). Figure~\ref{fig4}(a) shows the relative change of superfluid density and $T_{\rm c}$ at different applied pressures presented in a color contour plot. This plot allows us to directly visualize the influence of pressure on these superconducting parameters and make a comparison with other known superconductors\cite{GuguchiaMoTe2,Rohr,Castro, Khasanov3,Khasanov4, Khasanov5, Prando, GuguchiaFeAs} . Notably, in case of $\alpha$-PdBi$_2$, we observe a clear deviation from the horizontal line which accounts for the BCS superconductors. For comparison, we included few examples of unconventional superconductors which also demonstrate clear deviation from the horizontal line. For better visualization, we have also presented d$\sigma_{nor}$/d$p$ as a function of d$T_{\rm c, nor}$/d$p$ in Fig.~\ref{fig4}(b), which manifests that the data points corresponding to the unconventional superconductors situate away from the horizontal line. Moreover, we observed a linear dependence between $T_{\rm c}$ and the superfluid density for $\alpha$-PdBi$_2$ (purple dashed line in Fig.~\ref{fig4}(a)). Similar linear dependence was also observed in case of another TSC MoTe$_2$. In this context, it should be emphasised that the nearly linear relationship was originally observed in hole-doped cuprates in the underdoped region of the phase diagram \cite{Uemura,Uemura2}. Such linear relationship is expected only in the Bose-Einstein condensation (BEC)-like picture \cite{Uemura,Uemura3,Uemura4}. However, in case of cuprates the ratio between $T_{\rm c}$ and their effective Fermi temperature $T_{\rm F}$ is about $T_{\rm c}$/$T_{\rm F}\sim$~0.05, which suggests about four to five times reduction of $T_{\rm c}$ from the ideal BEC temperature for noninteracting Bose gas. These results were discussed in terms of the crossover from BEC to BCS-like condensation. Following the same line of reasoning, we anticipate similar mechanism in $\alpha$-PdBi$_2$.

In unconventional superconductors one of the major factors determining $T_{\rm c}$ is the superfluid density. Considering the linear dependency between $T_{\rm c}$ and superfluid density, we speculate that the suppression of $T_{\rm c}$ is a result of reduction of superfluid density in this compound. Another possibility to observe negative pressure effect on $T_{\rm c}$, is related to any weakening of the anharmonic electron-lattice interactions with increasing pressure \cite{Guguchia2}. However, this picture cannot explain the substantial ($\sim$20\%) reduction of superfluid density in $\alpha$-PdBi$_2$. Nevertheless, pressure dependent detailed density functional calculation will be crucial to address this conjecture.

The pressure effect on superconducting properties of $\alpha$-PdBi$_2$ reported by Zhou $et~al$.\cite{Zhou} was found to be irreversible. Therefore, it seemed very important to check whether the effect observed by us in $\alpha$-PdBi$_2$ were reversible or irreversible. We note that after releasing the pressure, we performed a quick temperature dependent TF-$\mu$SR measurements which revealed that ${\sigma}_{\rm sc} (T)$ overlaps with that observed before applying any pressure (for details, see SM \cite{supplemental}). This clearly suggests that the changes in superconducting properties seen here up to 1.77~GPa are completely reversible.

In summary, we have performed TF and ZF-${\mu}$SR measurements on a topological superconductor candidate $\alpha$-PdBi$_2$. TF-${\mu}$SR results reveal absence of any nodes in the superconducting gap structure of this compound and it can be best described by a fully gaped $s$-wave model. Observation of $s$-wave superconductivity in this system along with its unique band structure and Fermi surface topology, as demonstrated previously, endorse this compound as a promising candidate in search for Majorana zero modes. Further, our ZF-${\mu}$SR study manifests that time reversal symmetry is preserved in the superconducting state of $\alpha$-PdBi$_2$. Interestingly, ACS and TF-${\mu}$SR experiments under different hydrostatic pressures, reveal a negative pressure effect on $T_{\rm c}$ which is suppressed at an rate of -0.10~k/GPa, upon application of pressure up to 1.77~GPa. Notably, the superconducting gap value and the superfluid density also decrease with increasing pressure up to 1.77~GPa. Most intriguing observation of the present study is the linear dependence between $T_{\rm c}$ and superfluid density manifesting unconventional superconductivity in this compound. This also signals possible crossover from BEC to BCS in this compound. Further experimental and theoretical works under hydrostatic pressure condition are called for to fully comprehend the pressure dependent phase diagram of $\alpha$-PdBi$_2$. Our results on $\alpha$-PdBi$_2$ indicating a substantial content of novel physics in this compound opens more prospects for studies on TSC and its potential application.

\section{Acknowledgments}
The work was performed at the Swiss Muon Source (S${\mu}$S) Paul Scherrer Insitute, Villigen, Switzerland.  The work of R.G. was supported by the Swiss National Science Foundation (SNF Grant No. 200021-175935). DK was supported by the National Science Centre (Poland) under research grant no. 2015/18/A/ST3/00057. 

\end{document}


\preprint{preprint(\today)}

\title{Unconventional pressure dependence of the superfluid density in the nodeless  topological superconductor $\alpha$--PdBi$_2$}

\author{Debarchan Das}
\email{debarchandas.phy@gmail.com}
\affiliation{Laboratory for Muon Spin Spectroscopy, Paul Scherrer Institute, CH-5232 Villigen PSI, Switzerland}
\author{Ritu Gupta}
\affiliation{Laboratory for Muon Spin Spectroscopy, Paul Scherrer Institute, CH-5232 Villigen PSI, Switzerland}
\author{Christopher Baines}
\affiliation{Laboratory for Muon Spin Spectroscopy, Paul Scherrer Institute, CH-5232 Villigen PSI, Switzerland}
\author{Hubertus Luetkens}
\affiliation{Laboratory for Muon Spin Spectroscopy, Paul Scherrer Institute, CH-5232 Villigen PSI, Switzerland}
\author{Dariusz Kaczorowski}
\affiliation{Institute of Low Temperature and Structure Research, Polish Academy of Sciences, Wroc\l aw, ul. Ok\'olna 2, 50-422, Poland}
\author{Zurab Guguchia}
\affiliation{Laboratory for Muon Spin Spectroscopy, Paul Scherrer Institute, CH-5232 Villigen PSI, Switzerland}
\author{Rustem Khasanov}
\email{rustem.khasanov@psi.ch}
\affiliation{Laboratory for Muon Spin Spectroscopy, Paul Scherrer Institute, CH-5232 Villigen PSI, Switzerland}


\maketitle

\section{Supplemental Material }

\subsection{Sample preparation for $\mu$SR measurements}

\begin{figure}[htb!]
\includegraphics[width=1.0\linewidth]{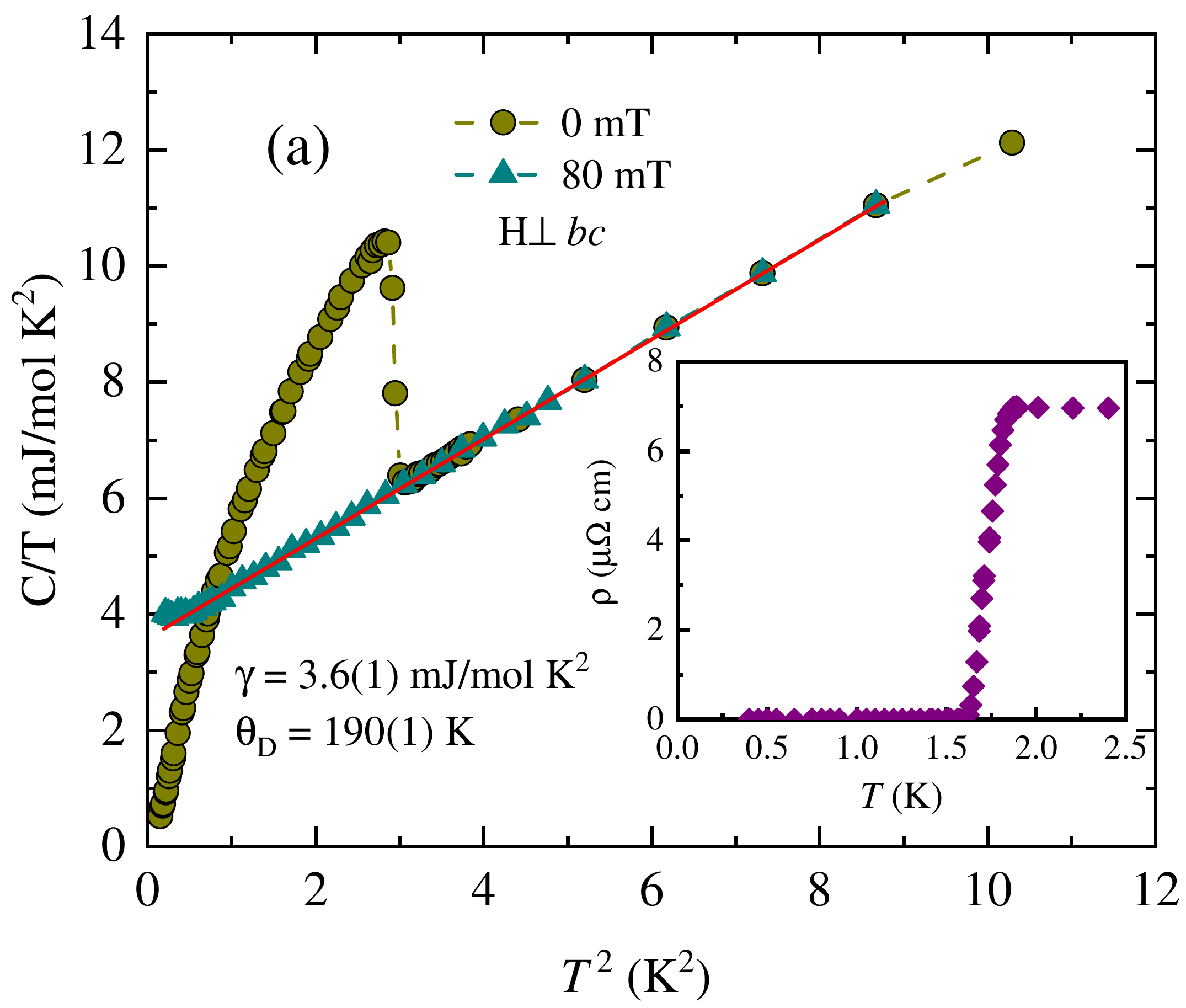}
\includegraphics[width=0.6\linewidth]{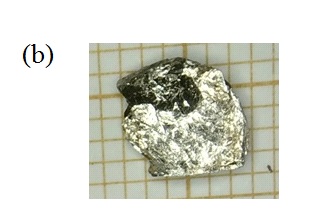}
\caption{(Color online) (a) C($T$)/$T$ ratio as a function of $T^2$ for zero field and applied field ($\bot$ $bc$ plane) of 80~mT.  Red solid line represents the fitting of the experimental data (under 80~mT) with C/$T$ = $\gamma_nT + \beta T^2$. Inset shows electrical resistivity as a function of temperature. (b) Photograph of one of the pieces of $\alpha$-PdBi$_2$ consisting of perhaps more than one crystals in it (not cleaved). While such a piece was used in measurements at the HAL-9500, many of such pieces were used for measurements at GPD.}
\label{figA1}
\end{figure}

In order to characterize the single crystals of $\alpha$-PdBi$_2$ , we performed preliminary investigation  using single crystal X-ray diffraction, electrical resistivity ($\rho$) and heat capacity ($C$) measurements. Fig~\ref{figA1}(a) shows the temperature dependence of the heat capacity, $C(T$), (presented as $C/T$ vs $T^2$) at zero and applied field of 80~mT. In the inset, we presented $\rho(T)$ highlighting the superconducting transition in this material. A prominent anomaly in  $C/T$ vs $T^2$ curve seen at $T_{\rm c}$ corroborates the bulk superconductivity in this compound. $C(T)$ data at 80~mT, can be well modeled by $C(T)/T = \gamma_{\rm n} + \beta T^2 $  (solid red line in Fig~\ref{figA1}(a)) yielding the normal state Sommerfeld coefficient $ \gamma_{\rm n} = 3.6(1)$~mJ/mol\,K$^2$ and the coefficient $\beta$ = 0.84(1)~mJ/mol\,K$^4$ which in turn provides an estimate of Debye temperature \cite{Kittel2005} $\Theta_{\rm D}$ = 190(1)~K. Using McMillan theory~\cite{McMillan}, we estimated the value of electron-phonon coupling parameter $\lambda_{\rm e-ph}$ = 0.54(2). A similar value was also seen in some other superconductors LaPt$_2$Si$_2$ ($\lambda_{\rm e-ph} = 0.53$), ThCoSi ($\lambda_{\rm e-ph} = 0.52$), 1T-PdTe$_2$ ($\lambda_{\rm e-ph} = 0.58$)~\cite{Das2, Domieracki, Anemone}. This value of $\lambda_{\rm e-ph}$ is indicative of weak electron-phonon coupling in $\alpha$-PdBi$_2$. In this context, it is interesting to note that $\lambda_{\rm e-ph}$ for $\alpha$-PdBi$_2$ is much smaller compared to that for $\beta$-PdBi$_2$ ($\lambda_{\rm e-ph} = 3.66$) determined using first principle calculation \cite{Sharma}.

One small plate like piece (5~mm$\times$6~mm$\times$1~mm) consisting of many single crystals (not cleaved) was mounted on a silver holder for measurements at HAL-9500 (see Fig~\ref{figA1}(b)). Magnetic field was applied perpendicular to the plane of this plate like piece. On the other hand, many such pieces were used for high pressure measurements on GPD using a double-wall piston cylinder type of cell with inner channel made of CuBe material. The filling factor of the pressure cell was maximized by accommodating as much sample as possible. The fraction of the muons stopping in the sample was approximately 45\%. We would like to emphasize that at room temperature, Daphne oil which was used as pressure transmitting medium, solidifies at a pressure of $\sim$2.3~ GPa \cite{Murata}. As the maximum pressure investigated in the present study was 1.74~GPa, we expect  hydrostatic conditions in the entire pressure range. The pressure was measured by tracking the superconducting transition of a very small indium piece placed inside the cell by ACS measurements in zero steady magnetic field with an ac-excitation voltage of 1.69~mV oscillating with frequency of 71.93~Hz.

\subsection{A\lowercase{nalysis of}  $\mu$SR \lowercase{data}}

At ambient pressure, below $T_{\rm c}$, $\mu$SR time-domain spectra were fitted using a three component expression with a Gaussian relaxation component
\begin{equation}
A_{TF}(t)= \sum\limits_{i=0}^3 A_i~e^{-\frac{\sigma_{i}^2t^2}{2}}\cos(\gamma_{\mu}B_it+\varphi),
\label{A1}
\end{equation}
here $A_i$, $\sigma_i$, and $B_i$ are the asymmetry, the relaxation rate and the mean field of the $i$th component, and $\varphi$ is the initial phase of the muon-spin ensemble. $\gamma_{\mu}/(2\pi)$ = 135.5342 MHz/T is the muon gyromagnetic ratio. Above $T_{\rm c}$, one component was sufficient to fit the spectra.
The first and second moments of the local magnetic field distribution, P(B), are given by \cite{Khasanov2}
\begin{equation}
\langle B \rangle= \sum\limits_{i=0}^3 \frac{A_i~B_i}{A_1+A_2+A_3},
\label{eq2}
\end{equation}

\begin{equation}
\langle \Delta B \rangle^2= \frac{\sigma^2}{\gamma_{\mu}^2}=\sum\limits_{i=0}^3 \frac{A_i}{A_1+A_2+A_3}[(\sigma_i/\gamma_\mu)^2+ (B_i-\langle B\rangle)^2],
\label{eq3}
\end{equation}
here, the superconducting part of the square root of the second moment ($\sigma_{\rm sc}$) was determined by subtracting the nuclear moment contribution ($\sigma_{\rm nm}$) measured at $T>T_{\rm c}$ using $\sigma_{\rm sc}^2$ =$\sqrt{\sigma^2-\sigma_{\rm nm}^2}$.
\begin{figure}[htb!]
\includegraphics[width=1.0\linewidth]{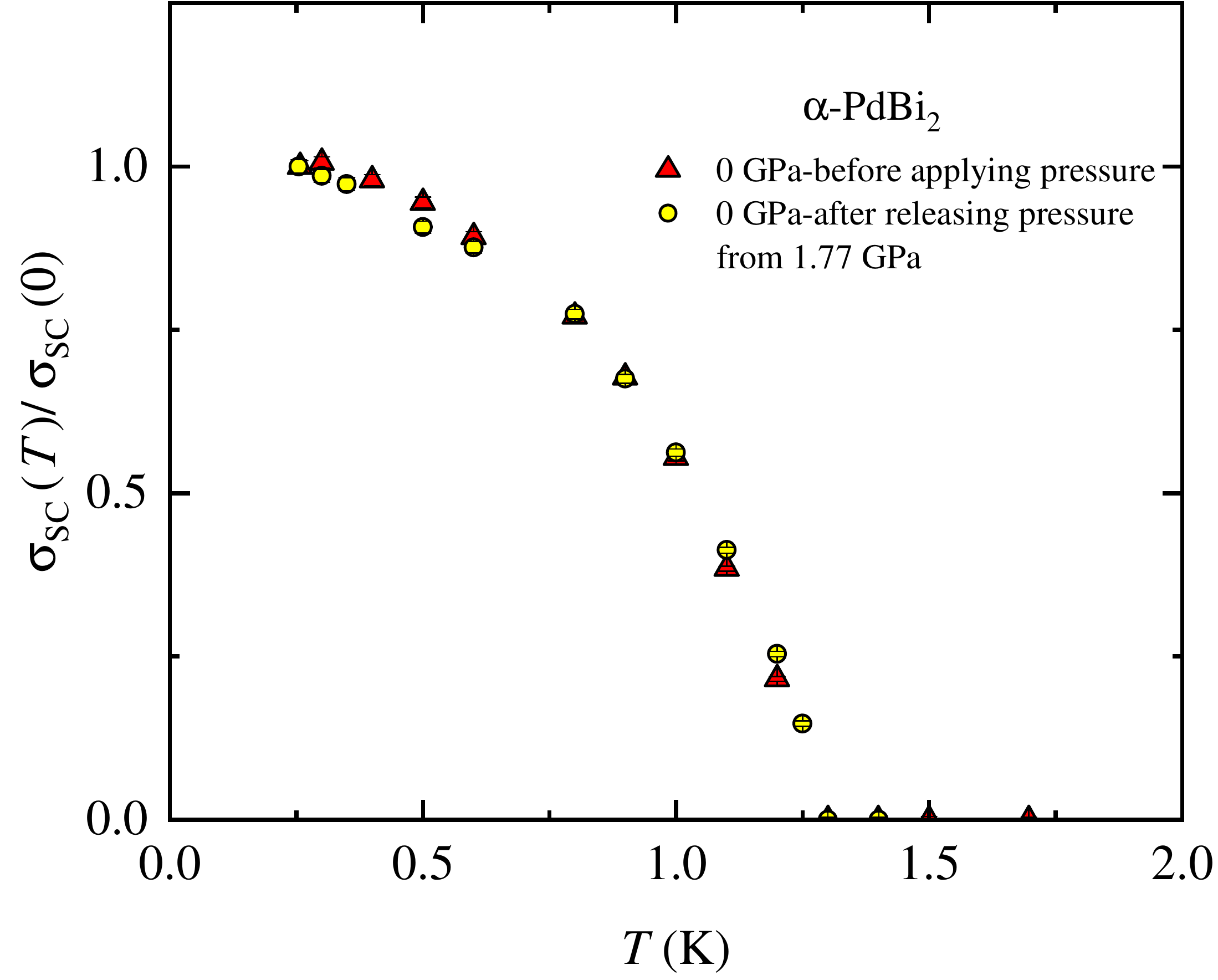}
\caption{(Color online) The temperature dependence of ${\sigma}_{\rm sc} (T)$ , normalized to its value for the zero applied pressure, before applying pressure and after releasing pressure (from 1.77~GPa).}
\label{figA2}
\end{figure}

Within the London model, ${\sigma}_{\rm sc}(T)$ can be modeled by the following expression:\cite{Suter,Tinkham,Prozorov}
\begin{equation}
\frac{{\sigma}_{\rm sc}(T)}{{\sigma}_{\rm sc}(0)}=
1+\frac{1}{\pi}\int_{0}^{2\pi}\int_{\Delta(_{T,\varphi})}^{\infty}\left(\frac{\partial f}{\partial E}\right)\frac{EdEd\varphi}{\sqrt{E^2-\Delta_i(T,\varphi)^2}},
\label{sig}
\end{equation}
where $f=[1+\exp(E/k_{\rm B}T)]^{-1}$ is the Fermi function, ${\varphi}$ is the azimuthal angle along the Fermi surface, and ${\Delta}_{i}(T,{\varphi})={\Delta}_{0,i}{\Gamma}(T/T_{\rm c})g({\varphi}$) (${\Delta}_{0,i}$ is the maximum gap value at $T=0$). The temperature dependence of the gap is described by the expression ${\Gamma}(T/T_{\rm c})=\tanh{\{}1.82[1.018(T_{\rm c}/T-1)]^{0.51}{\}}$,\cite{carrington}. Here, $g({\varphi}$) corresponds to the angular dependence of the gap and  takes the value 1 for $s$-wave and ${\mid}\cos(2{\varphi}){\mid}$ for a $d$-wave gap \cite{Fang} resulting in nodal points for  ${\varphi}$ = $\pm \pi/4$, $\pm 3\pi/4$. Solid lines through the data points in Fig. 1(c) of the main manuscript represents fitting of the experimental data with these models.

To analyze the $\mu$SR asymmetry spectra obtained from GPD experiments, we used following functional form:

\begin{multline}
A_{TF}(t)= A_S~e^{-\frac{\sigma^2t^2}{2}}\cos(\gamma_{\mu}B_{int}t \varphi) \\+ A_{pc}~e^{-\frac{\sigma_{pc}^2t^2}{2}}\cos(\gamma_{\mu}B_{ext}t+\varphi)  ,
\label{eq1}
\end{multline}
where $A_S$ and $A_{pc}$ represent the initial asymmetries of the $\mu$SR signal for the sample and the pressure cell, respectively. and $\varphi$ is the initial phase of the muon-spin ensemble. $B_{int,s}$ and $B_{ext}$ are the internal magnetic fields at the muon site in the sample and external applied field, respectively.  $\sigma_{pc}\simeq~0.28~\mu s^{-1}$, is the field and the temperature independent relaxation rate of beryllium-copper. The fits were performed globally. $\sigma_{sc}$ was extracted from $\sigma$ adapting the same formalism used for ambient pressure measurements. 

\subsection{Reversibility check of pressure effects in $\alpha$-PdBi$_2$}

In the previous high pressure study, Zhou $et~al$.\cite{Zhou} observed clear irreversible pressure induced effects in $\alpha$-PdBi$_2$ while releasing pressure from 42.9~GPa. Pressure dependent structural investigation suggested that there is a transformation from alpha to beta phase at 8~GPa. Even though in our present study the  highest applied was 1.77~GPa which is much smaller than in ref.~\cite{Zhou}, it was essential to check whether the pressure induced changes of superconducting properties observed by us, were reversible or irreversible. In order to address this question, we measured temperature dependent TF-$\mu$SR measurements after releasing the pressure to 0~GPa from 1.77~GPa keeping the sample in the pressure cell. As can be seen from Fig.~\ref{figA2}, $\sigma_{\rm SC}(T)$ data obtained before closing and after releasing the pressure overlaps with each other evidencing that the pressure induced (up to 1.77~GPa) changes of superconducting properties in $\alpha$-PdBi$_2$ reported in this paper, are reversible.